\documentclass[sigconf,authorversion,nonacm]{acmart}

\usepackage[table]{xcolor}
\usepackage{algorithmic}
\usepackage{textcomp}
\usepackage{comment}
\usepackage{makecell}
\usepackage{listings}
\usepackage{adjustbox}
\usepackage{threeparttable}

\usepackage{pifont}

\copyrightyear{2022}
\acmYear{2022}
\setcopyright{usgovmixed}\acmConference[CSET 2022]{Cyber Security Experimentation and Test Workshop}{August 8, 2022}{Virtual, CA, USA}
\acmBooktitle{Cyber Security Experimentation and Test Workshop (CSET 2022), August 8, 2022, Virtual, CA, USA}
\acmPrice{15.00}
\acmDOI{10.1145/3546096.3546112}
\acmISBN{978-1-4503-9684-4/22/08}

\begin{document}

\title{A Broad Comparative Evaluation of x86-64 Binary Rewriters}

\author{Eric Schulte}
\affiliation{%
  \institution{}
  \city{}
  \country{}}
\email{schulte.eric@gmail.com}

\author{Michael D. Brown}
\affiliation{%
  \institution{Trail of Bits}
  \city{New York, NY}
  \country{USA}}
\email{michael.brown@trailofbits.com}

\author{Vlad Folts}
\affiliation{%
  \institution{Grammatech, Inc.}
  \city{Ithaca, NY}
  \country{USA}}
\email{vfolts@grammatech.com}

\renewcommand{\shortauthors}{Schulte, et al.}

\begin{abstract}
Binary rewriting is a rapidly-maturing technique for modifying software 
for instrumentation, customization, optimization, and hardening without 
access to source code. Unfortunately, the practical applications of binary 
rewriting tools are often unclear to users because their limitations are
glossed over in the literature. This, among other challenges, has prohibited the
widespread adoption of these tools. To address this shortcoming, we
collect ten popular binary rewriters and assess their
\textit{generality} across a broad range of input binary classes and 
the functional \textit{reliability} of the
resulting rewritten binaries. Additionally, we evaluate the \textit{performance}
of the rewriters themselves as well as the rewritten binaries 
they produce. 

The goal of this broad evaluation is to establish a shared context for future 
research and development of binary rewriting tools by providing a \textit{state of the
practice} for their capabilities. To support potential binary rewriter users, 
we also identify input binary features that are predictive of tool success 
and show that a simple decision tree model can accurately predict 
whether a particular tool can rewrite a target binary. The binary rewriters, 
our corpus of 3344 sample binaries, and the evaluation infrastructure itself 
are all freely available as open-source software.

\end{abstract}



\maketitle

\section{Introduction}

Binary rewriting (also referred to as recompilation) is an emerging 
research area that has been enabled by recent advances in binary 
analysis. Binary rewriting tools have the potential to address 
long-standing problems in cyber security by enabling binary analysis, 
patching, and security hardening for programs where source code is 
not available (e.g., legacy or closed-source software). For binary 
rewriting tools to be viable, they must \textit{generalize}
to the full variety of programs available on heterogeneous computing
platforms and \textit{reliably} produce functional rewritten
binaries. 

A surfeit of research into binary rewriting applications including
instrumentation, optimization, configuration, debloating, and
hardening reveals a wide and largely under-emphasized variance in the
generality and reliability of binary rewriters
\cite{romer1997instrumentation, PSI, pebil, williams2020egalito,
  dinesh2019retrowrite, schwarz2001plto, diablo, tilevich2005binary,
  qian2019razor, bincfi, secondwrite, reins, duck2020binary,
  di2017rev}. Frequently, papers presenting these tools only briefly
mention large gaps in generality such as support limited to binaries
with relocation information and/or symbols -- neither are typical of
commercial off-the-shelf (COTS) software~\cite{williams2020egalito,
  dinesh2019retrowrite}.

While there has been significant research to date seeking to
systematize knowledge of general binary rewriting
techniques~\cite{wenzl2019hack} and evaluate the quality of binary
lifters and disassemblers~\cite{meng2016binary, andriesse2016}, no
systematic comparative evaluation of binary rewriters has yet
been conducted. In this work, we address this important knowledge gap
by conducting such an evaluation of 10 binary rewriters across 3344
variant binaries sourced from 34 benchmark programs and 3 production
compilers. The tools evaluated in this work are \textit{static} binary
rewriting tools; we exclude dynamic binary
rewriting tools (e.g., PIN ~\cite{luk2005pin})
whose runtime harnesses and overhead often make them impractical for
the applications considered by this work.

Our work differs from previous surveys in two key ways. First, prior
work systematizing knowledge on binary rewriters~\cite{wenzl2019hack}
focused primarily on their underlying techniques and algorithms and as
such did not evaluate their artifacts empirically. In contrast, our
evaluation focuses on measuring and comparing the {\em generality} and
{\em reliability} of a broad collection of publicly available binary
lifter and rewriter tools.\footnote{We are not aware of any comparable
closed source binary rewriters.} Second, prior works performing
comparative evaluation of binary disassmblers and lifters
\cite{meng2016binary, andriesse2016} focus on \textit{depth} achieving
near complete measurement of binary analysis accuracy across a small
pool of binaries. The difficulties implicit in a truly thorough
analysis limits the breadth of these works to small numbers of
binaries or to specific classes of binaries. Our evaluation focuses on
input program \textit{breadth} to directly address tool
\textit{generality} and rewritten binary functionality to directly
address \textit{reliability}.

\textbf{Summary of Contributions.} In this paper, we first review
related work evaluating binary transformation tools in
Section \ref{sec:background}. We then describe our experimental
methodology to assess the generality and reliability of existing tools
in Section \ref{sec:methodology}. Next, we present our experimental
results and predictive models derived from them in Section 
\ref{sec:results}. Finally, we discuss the state of practice in 
binary rewriting tools and options for potential users in Section
\ref{sec:conclusion}.



\section{Background}
\label{sec:background}

\subsection{Types of Binary Rewriters}

\textbf{Direct Rewriting.}  Zipr~\cite{hiser2014framework} lifts a
binary into an Intermediate Representation Data Base
(IRDB~\cite{irdb}) upon which transformations may be applied. The IRDB
can then be written directly to a binary executable.  Similarly,
Egalito~\cite{williams2020egalito} lifts a binary into a low-level
Intermediate Representation (IR), provides an API for transforming
this IR, and supports lowering this IR back to a modified binary.

\textbf{Reassemblable Disassemblers.} Uroboros~\cite{uroboros} --
superseded by the Phoenix reassemblable decompiler~\cite{phoenix} --
popularized the idea of reassemblable disassembly (i.e., disassembly
to assembly code that can be readily reassembled).
Retrowrite~\cite{dinesh2019retrowrite} also emits reassemblable
assembly code.
DDisasm~\cite{ddisasm} lifts to GrammaTech's Intermediate
Representation for Binaries (GTIRB)~\cite{gtirb}, and can also
directly emit assembly code that can be reassembled.

\textbf{LLVM Rewriting.}  The now defunct SecondWrite~\cite{secondwrite}
was the first tool to lift binaries to LLVM
IR. More recently, McSema~\cite{mcsema} and rev.ng~\cite{di2017rev} have become the predominant
binary lifters to target LLVM IR. Although LLVM has a large user
community and provides many analysis, optimization, and hardening
passes, there are two properties of its IR that make it difficult for
lifters to target. First, it requires \textit{type} information to
represent data.  This requires binary type analysis, which is 
prohibitively difficult at the scale required to rewrite large programs.
Instead, many tools explicitly \textit{emulate} the stack, stack 
maintenance around function calls, and memory as a large array of bytes.
The lack of true types and stack/memory emulation limits the 
utility of most existing LLVM passes and 
results in baroque and inefficient rewritten binaries~\cite{kiaei2020rewrite}. Second, LLVM 
represents code in static single assignment form resulting in a 
loss of specificity with respect to the original 
instructions.

\textbf{Trampoline Rewriting.} Trampoline rewriters such as e9patch minimally disturb the memory
image of the original binary~\cite{Dyninst, duck2020binary}.  New code is placed outside of the original image
and the only changes within the image are jumps to this new code, which itself
jumps back to the original code.  Trampoline rewriting can be very robust as it requires minimal binary analysis
(e.g., symbolization is not necessary), however it is only well-suited to {\em additive} transformation. The original 
code cannot be modified, moved, optimized, or removed easily or effectively.

\subsection{Related Work}

Wenzl et. al.~\cite{wenzl2019hack} survey over 50 years of binary rewriter research with the primary objective of categorizing binary rewriting approaches by end use-case, analysis techniques, code transformation methods, and code generation methods. However, this survey does not include a comparative evaluation of the tools presented in the literature. Our work aims to complement and extend their survey by providing an empirical evaluation of binary rewriters.

Several works performing extensive evaluations of binary disassemblers have been published in recent years \cite{andriesse2016, meng2016binary, li2020generation, pang2021sok}. Collectively, these works thoroughly document the approaches, challenges, trade-offs, and shortcomings of disassemblers. Further, they establish that modern disassemblers achieve high accuracy (close to 100\%) even in the presence of challenging (although rare) code constructs due to advances and specialization in code discovery and control-flow reconstruction algorithms. These evaluations are similar in scale to our work, however we evaluate tools with custom disassembly routines not covered in these works (all except uroboros). Further, our evaluation focuses on binary rewriting as opposed to disassembly, which is a prerequisite step to rewriting.

Additionally, several evaluation frameworks and benchmark suites for evaluating binary analysis tools related to binary rewriting have been recently proposed. Mishegos~\cite{woodruff21differential} is a novel differential fuzzer that can be used to evaluate x86-64 instruction decoders. ConFIRM~\cite{xu2019confirm} is an evaluation framework and benchmark suite for assessing software hardening transformations, namely control-flow integrity implementations. Finally, Dasgupta et. al.~\cite{10.1145/3385412.3385964} recently presented a scalable technique for validating and detecting bugs in binary lifters such as McSema~\cite{mcsema}.

\section{Methodology}
\label{sec:methodology}

\subsection{Tool Selection}

We selected ten binary rewriters for our evaluation, listed in Table
\ref{tab:tools}. While our corpus of tools is not exhaustive, it
provides excellent coverage of tools that are mature, robust, and scale 
via automation. We exclude two notable binary rewriting tools, McSema\cite{mcsema} 
and Ramblr\cite{ramblr}, from our evaluation.

In McSema's case, the tool can be automated to lift a wide variety of binaries to 
LLVM IR. In our initial evaluation, McSema successfully lifted 57\% of the 3344 program variants
we tested it against. However, McSema's rewriting workflow currently requires manual steps by 
a knowledgeable operator. As a result, rewriting binaries with McSema does not satisfy the
requirements of our evaluation. While still maintained, McSema's shortcomings in this area are largely
due to age (its last major release was in 2018) and the recent modernization of its dependent 
libraries, Anvill~\cite{anvill} and Remill~\cite{remill}, to support newer lifting tools such 
as Rellic~\cite{rellic}.

In Ramblr's case, it is implemented as a supplementary analysis for the Angr binary analysis 
framework~\cite{shoshitaishvili2016state}. In its current from, Ramblr is capable of reassembling 
x86 and x86\_64 binaries disassembled by Angr; however, it does not expose an interface for 
transforming disassembled binaries. As such, Ramblr cannot be truly used as a binary rewriter 
(i.e., it cannot perform our second evaluation task) "out of the box".


\begin{table}[!htb]
    \centering
    \caption{Tools selected for this evaluation}
    \label{tab:tools}
    {\rowcolors{2}{lightgray!50}{white}
\begin{tabular}{|l|c|}
    \hline
    \textbf{Tool} & \textbf{Rewriter Type} \\ 
    \hline
    ddisasm \cite{ddisasm}&    Reassemblable Disassembler \\ 
    e9patch \cite{duck2020binary}  & LLVM Trampoline \\ 
    Egalito \cite{williams2020egalito}  & Direct Rewriter \\ 
    mctoll \cite{mctoll}   & LLVM Rewriter \\ 
    multiverse \cite{superset} & Direct Rewriter \\ 
    ReOpt  \cite{reopt} & LLVM Rewriter \\ 
    revng   \cite{di2017rev} & LLVM Rewriter \\ 
    
    Retrowrite \cite{dinesh2019retrowrite} & Reassemblable Disassembler  \\ 
   
    Uroboros \cite{uroboros} &   Reassemblable Disassembler \\ 

    Zipr \cite{hiser2014framework} &      Direct Rewriter \\ 
    
    \hline
\end{tabular}
}

\end{table}

\subsection{Evaluation Variant Generation}

In order to obtain realistic evaluation results, we combined benchmark
lists compiled by two program managers from the United States
Department of Defense to arrive at a diverse list of 34 real-world
benchmark programs, shown in Table \ref{tab:benchmarks}. To this corpus we also 
added a trivial ``Hello World!'' program to our
corpus to provide a low-water mark for program complexity. For each benchmark and the {\tt hello-world} program, 
we compiled an x86-64 variant of the program using one permutation of the compilers, optimization levels, code layout, symbol options, and operating systems listed in Table \ref{tab:perms}. In total, we generated 3344 distinct evaluation variants.

\begin{table*}[!htb]
    \centering
    \caption{Benchmark programs used in this evaluation}
    \label{tab:benchmarks}
    {\rowcolors{2}{lightgray!50}{white}
\begin{tabular}{|c|r|l|c|r|l|}
\hline
Program & SLOC & Description & Program & SLOC & Description \\
\hline
\href{https://github.com/anope/anope}{anope}                                  & 65,441    & IRC Services             & \href{https://openvpn.net/}{openvpn}                                          & 89,312    & VPN Client               \\
\href{https://www.asterisk.org/}{asterisk}                                    & 771,247   & Communication Framework  & \href{https://developer.pidgin.im/wiki/UsingPidginMercurial}{pidgin}          & 259,398   & Chat Client              \\
\href{https://www.isc.org/bind/}{bind}                                        & 376,147   & DNS System               & \href{http://pks.sourceforge.net/}{pks}                                       & 40,788    & Public Key Server        \\
\href{https://en.bitcoin.it/wiki/Bitcoind}{bitcoind}                          & 229,928   & Bitcoin Client           & \href{https://poppler.freedesktop.org}{poppler}                               & 188,156   & PDF Reader               \\
\href{http://www.thekelleys.org.uk/dnsmasq/doc.html}{dnsmasq}                 & 34,671    & Network Services         & \href{http://mirror.jaleco.com/postfix-release/}{postfix}                     & 134,957   & Mail Server              \\
\href{https://filezilla-project.org/ (FTP server)}{filezilla}                 & 176,324   & FTP Client and Server    & \href{http://www.proftpd.org/}{proftpd}                                       & 544,178   & FTP Server               \\
\href{https://gitlab.gnome.org/GNOME/gnome-calculator}{gnome-calculator}      & 301       & Calculator               & \href{http://cr.yp.to/qmail.html}{qmail}                                      & 14,685    & Message Transfer Agent   \\
\href{http://leafnode.sourceforge.net/download.shtml}{leafnode}               & 12,945    & NNTP Proxy               & \href{https://redis.io/}{redis}                                               & 14,685    & In-memory Data Store     \\
\href{https://github.com/LibreOffice/core}{Libreoffice}                       & 5,090,852 & Office Suite               & \href{https://www.samba.org/}{samba}                                          & 1,863,980 & Windows Interoperability \\
\href{https://github.com/zeromq/libzmq}{libzmq}                               & 62,442    & Messaging Library             & \href{https://www.proofpoint.com/us/open-source-email-solution}{sendmail}     & 104,450   & Mail Server              \\
\href{https://www.lighttpd.net/}{lighttpd}                                    & 89,668    & Web Server        & \href{http://www.gnu.org/software/sipwitch/}{sipwitch}                        & 17,134    & VoIP Server              \\
\href{https://github.com/memcached/memcached}{memcached}                      & 33,533    & In-memory Object Cache               & \href{https://www.snort.org/}{snort}                                          & 344,877   & Intrusion Prevention     \\
\href{https://getmonero.org/resources/user-guides/vps_run_node.html}{monerod} & 394,783   & Blockchain Daemon         & \href{https://sqlite.org/index.html}{sqlite}                                  & 292,398   & SQL Server               \\
\href{https://mosh.org/}{mosh}                                                & 12,890    & Mobile Shell  & \href{http://www.squid-cache.org/}{squid}                                     & 212,848   & Caching Web Proxy        \\
\href{https://github.com/mysql/mysql-server}{mysql}                           & 3,331,683 & SQL Server                   & \href{https://www.unrealircd.org/}{unrealircd}                                & 90,988    & IRC Server               \\
\href{https://nginx.org/en/}{nginx}                                           & 170,602   & Web Server          & \href{https://github.com/vim/vim}{vi/vim}                                     & 394,056   & Text Editor              \\
\href{https://github.com/openssh/openssh-portable}{ssh}                       & 127,363   & SSH Client and Server                & \href{http://infozip.sourceforge.net/Zip.html#Sources}{zip}                   & 54,390    & Compression Utility      \\

\hline
\end{tabular}
}

\end{table*}

\begin{table}[htb!]
    \centering
    \caption{Variant configuration options}
    \label{tab:perms}
    {\rowcolors{2}{lightgray!50}{white}
\begin{threeparttable}
\begin{tabular}{|c|c|c|c|c|}
    \hline
    \textbf{Compiler} & \textbf{Flags}  & \makecell{\textbf{Relocation} \\ (Position-)} & \textbf{Symbols} & \makecell{\textbf{Operating} \\ \textbf{Systems}} \\
    \hline
    clang & O0 & Independent & Present & Ubuntu 16.04$^{1}$ \\
    gcc & O1 & Dependent & Stripped & Ubuntu 20.04 \\
    icx & O2 & & & \\
    & O3 & & & \\
    & Os & & & \\
    & Ofast & & & \\
    \hline 
    OLLVM & fla & Independent & Present & Ubuntu 20.04 \\
    & sub & Dependent & Stripped & \\
    & bcf$^{2}$ & & & \\
    \hline
\end{tabular}
\begin{tablenotes}
      \scriptsize
      \item $^{1}$Some binaries could not be built on this OS due to unavailable dependencies.
      \item $^{2}$Probability variable set to always insert (100\%)
    \end{tablenotes}
\end{threeparttable}
}
\end{table}

\subsection{Evaluation Tasks}
\label{sec:eval-tasks}

We evaluate our selected binary rewriters based on their ability to
successfully perform two rewriting {\em tasks} and record their
progress at multiple {\em checkpoints}. In the interest of breadth we
use a proxy for successful (i.e., correct) rewriting. Specifically, we
consider a rewrite to be successful if the output executable passes a
very simple functional test (described in \autoref{sec:evalmetrics}). In practice many binary rewriting tools fail
fast when problems arise, meaning they either completely fail to
produce a new executable or they produce a executable that is unable
to start execution.  For a small subset of our benchmark with readily
executable test suites with high coverage we also tested programs
against the full test suite.

\textbf{Tasks.} The two tasks we use to evaluate our tools are:

\begin{description}
\item[NOP] This task is a minimal NOP (i.e., No Operation) transform
  that simply lifts the binary and then rewrites without
  modification. The NOP transform tests the ability of a binary
  rewriter to successfully process the input binary, populate its
  internal or external intermediate representation of the binary, and
  then produce a new rewritten executable. Despite its name this
  transform is decidedly non-trivial for most rewriters, evidenced by
  the fact that rewritten binaries typically look very different from
  the original.
\item[AFL] This task is a more complex transform characteristic of the
  needs of instrumentation, e.g. to support gray-box fuzz testing.  It
  evaluates our tools' abilities to extensively {\em transform} a binary
  with instrumentation to support AFL++~\cite{aflpp}\footnote{Each
  tool we selected except multiverse claims to support AFL++'s
  instrumentation.}.  This task is important to include as many
  rewriters cover up analysis errors by incorporating reasonable
  defaults (e.g., linking code from the original binary on lifting
  failure, or preserving unidentified symbols which continue to
  resolve correctly if code is left undisturbed in memory).
\end{description}

\textbf{Checkpoints.} For every attempted rewrite operation, we collect some of the following artifacts to checkpoint the process:
\begin{description}
\item[IR] For every binary rewriting tool that leverages some form of
  external IR, we collect that IR.  Specifically, we collect the ASM
  files generated by tools that emit reassemblable disassembly and the
  LLVM IR for tools targeting LLVM.  Zipr, Egalito, and multiverse use
  a internal IRs that are not easily serialized to disk.  E9patch does
  not present the original code for rewriting.  As such, we do not
  track successful IR generation for these tools.
\item[EXE] We next check if the rewriter successfully creates a new
  executable. In some cases rewritten executables are 
  trivially recreated and are not an indicator of success (e.g., Egalito 
  almost always generates a new executable even if most of them are 
  non-functional). However, in most cases
  the ability to re-assemble and re-link a new executable indicates
  that the rewriting tool both successfully disassembled reasonable
  assembly instructions and generated the required combinations of
  symbols and included libraries.
\end{description}

\subsection{Evaluation Metrics}
\label{sec:evalmetrics}

\textbf{Functional Metrics.} To measure rewriter correctness, we first
observe the success rate for each tool across all variants for
both tasks (i.e., NOP and AFL) at both checkpoints (i.e., IR and
EXE). Next, we perform a simple invocation of the NOP rewritten
programs (e.g., running {\tt -{}-help}) to ensure the
rewrite did not obviously corrupt the program. We refer to this
as the \textit{Null Function} test. Finally, we execute the AFL
rewritten programs with the driver provided by AFL++ to ensure
instrumenting the program via binary rewriting did not corrupt the
program and that instrumentation was successfully incorporated.  We
refer to this as the \textit{AFL Function} test.

\textbf{Non-Functional Metrics.} To measure rewriter runtime
performance we observe the total required runtime and the memory
high-water mark used by tools during rewriting. Available memory is
often the limiting factor for rewriting because many underlying
analyses scale super-linearly in the input binary size.

To determine the performance impacts of rewriting on binaries, we
first measure file size impacts using Bloaty~\cite{bloaty}. Size is an
important metric as it measures the degree to which a rewriting tool
has inserted dynamic emulation or runtime supports -- with their
associated increased complexity and runtime costs. Finally, for
successfully rewritten program variants with publicly available test
suites we measure the impact of rewriting on
performance. Specifically, we measure pass rate for all tests in the
suite, runtime of the test suite, and the memory consumption
high-water mark during execution of the full test suite as compared to
the original.

\section{Experimental Results}
\label{sec:results}

We present binary rewriter success both in aggregate across our entire
benchmark set and broken out into cohorts. Each cohort of binaries has like
characteristics that highlight the comparative strengths and weaknesses
of the evaluated tools.

\subsection{Aggregate Rewriting Success Rates}

\begin{table*}[!htb]
     \begin{minipage}{\textwidth}
\centering
\caption{Number and percentage of x86-64 Linux binaries for
which the rewriter successfully produces IR, produces a NOP-transformed executable (``EXE''), 
passes the Null Function test, produces an AFL++ instrumented 
executable (``AFL EXE''), and passes the AFL Function test. Data is aggregated across the the full suite of binaries (3344) in the first set of columns and across position-independent, non-stripped binaries (1672) in the second set.}
    \label{tab:aggregate_success}
      \makeatletter
\newcommand\footnoteref[1]{\protected@xdef\@thefnmark{$#1$}\@footnotemark}
\makeatother
{\rowcolors{4}{lightgray!50}{white}
\begin{tabular}{|c|c|c|c|c|c||c|c|c|c|c|c|}
\cline{2-11}
\multicolumn{1}{c|}{}& \multicolumn{5}{c||}{\textbf{Full Suite}} & \multicolumn{5}{c|}{\textbf{Position-Independent, Symbols Present only}} \\
\hline
Tool       & IR      & EXE      & \makecell{Null \\Func.}   & AFL   EXE          & \makecell{AFL \\         Func.}     & IR       & EXE     & \makecell{Null \\Func.} & AFL EXE & \makecell{AFL \\ Func.} \\
\hline
ddisasm    & 3282    & 2972     & 2873     & 3020     & 2346    & 1638    & 1428     & 1385     & 1456     & 1125    \\
\%         & 98.14\% & 88.87\%  & 85.91\%  & 90.31\%  & 70.15\% & 97.96\% & 85.40\% & 82.82\% & 87.08\% & 67.28\% \\
e9patch    & NA      & 3344     & 2620     & 3344     & 1212    & NA      & 1672     & 1310     & 1672     & 607     \\
\%         & NA      & 100\% & 78.34\%  & 100\% & 36.24\% & NA      & 100\% & 78.34\% & 100\% & 36.30\% \\
egalito    & NA      & 3294     & 983      & 2493     & 0       & NA      & 1661     & 492      & 1261     & 0       \\
\%         & NA      & 98.50\%  & 29.39\%  & 74.55\%  & 0       & NA      & 99.34\% & 29.42\% & 75.40\% & 0       \\
mctoll     & 30      & 30       & 30       & 30       & 30      & 30      & 30       & 30       & 30       & 30      \\
\%         & .89\%   & .89\%    & .89\%    & .89\%    & .89\%   & 1.78\%  & 1.78\%   & 1.78\%   & 1.78\%   & 1.78\%   \\
multiverse & NA      & 880      & 362      & 0        & 0       & NA      & 437      & 181      & 0        & 0       \\
\%         & NA      & 26.31\%  & 10.82\%  & 0        & 0       & NA      & 26.12\% & 10.82\%  & 0       & 0       \\
reopt      & 3007    & 2556     & 1134     & 0        & 0       & 1504    & 1425     & 684      & 0        & 0       \\
\%         & 89.92\% & 76.43\%  & 33.91\%  & 0        & 0       & 89.94\% & 85.22\% & 40.90\% & 0       & 0       \\
retrowrite & 817     & 334      & 309      & 330      & 254     & 817     & 334      & 309      & 330      & 254     \\
\%         & 24.43\% & 9.98\%   & 9.24\%   & 9.86\%   & 7.59\%  & 48.86\% & 19.96\%  & 18.48\%  & 19.72\%  & 15.18\%  \\
revng      & NA\footnote{Although rev.ng {\em can} produce IR, its normal usage is to output rewritten binaries directly which is how it was run in this evaluation.}      & 885      & 786      & 0        & 0       & NA\footnoteref{a}      & 439      & 390      & 0        & 0       \\
\%         & NA\footnoteref{a}      & 26.46\%  & 23.50\%  & 0        & 0       & NA\footnoteref{a}      & 26.26\% & 23.32\% & 0       & 0       \\
uroboros   & 364     & 216      & 96       & 210      & 0       & 183     & 108      & 48       & 105      & 0       \\
\%         & 10.88\% & 6.45\%   & 2.87\%   & 6.27\%   & 0       & 5.47\%  & 6.44\%  & 2.86\%  & 6.26\%  & 0       \\
zipr       & NA      & 3344     & 3344     & 2708     & 1609    & NA      & 1672     & 1672     & 1560     & 1079    \\
\%         & NA      & 100\% & 100\% & 80.98\%  & 48.11\% & NA      & 100\% & 100\% & 93.30\% & 64.52\% \\
\hline
\end{tabular}
}

     \end{minipage}
\end{table*}

Our aggregate success results are presented in
\autoref{tab:aggregate_success}. Overall, we observed a very broad range of success rates (and by extension levels of support) 
achieved by our selected binary rewriting tools. For the NOP 
transform, the tools fall into four distinct categories characterized by the fraction of the universe of potential binaries
they can handle:

\begin{enumerate}
    \item Tools that work only on a tiny fraction ($\le$5\%) of binaries. 
    This group includes mctoll and uroboros.
    \item Tools which work on a few ($\sim$10\%) binaries. 
    This group includes multiverse and retrowrite.
    \item Tools that work on some ($\sim$33\%) binaries. 
    This group includes egalito, reopt, and revng.
    \item Tools that work on most ($\ge$80\%) binaries.
    This group includes ddisasm, e9patch, and zipr.
\end{enumerate}

In category (1) we find tools that handle a very limited set
of binaries (e.g., mctoll only successfully transformed {\tt
  hello-world} binaries).  The tools in category (2) support a wider
range of binaries but in many cases make hard
and fast assumptions (e.g., multiverse can only successfully rewrite 
binaries compiled by old versions of clang and gcc available on Ubuntu 16.04).  
These tools still do not handle binaries that make use of fairly common code structures (e.g., C++
exceptions).  The tools in category (3) largely only work with
relocation and debug information,%
\footnote{The exception is rev.ng which {\em can} work without relocation or debug information.}
but are able to handle a wide range
of the binaries meeting these restrictions.  Finally, category (4)
tools do {\em not} require relocation or debug information and support
a wide range of both complex code structures and compiler-specific
binary generation behaviors such as multiple forms of jump tables,
and data intermixed with code.

Given that so many tools require relocation and debug information we
present a second view of our results limited to binaries that include
this information (i.e., non-stripped, position-independent variants)
in the right of \autoref{tab:aggregate_success}. Although
position-independent binaries are increasingly common as ASLR becomes
the norm, it is still uncommon for COTS binaries to include
debug information.

These results show the increase in
rewriting success rate that a developer might expect if they compile 
their binaries with relocation and debug information to support binary 
rewriting.
However, such binaries are not characteristic of the stripped COTS binaries likely
to be received from third parties or found in the wild. 

As shown in \autoref{tab:aggregate_success}, the AFL
transform provides a much better proxy for actual binary rewriter 
performance than the NOP transform. This is true for at least two
reasons.  First, when applying the NOP transform many relative and
absolute locations in a rewritten binary will continue to match their
locations in the original binary because no attempt is made to modify
the lifted code.  This provides a great deal of grace for rewriters
that missed code or symbols in the original binary because symbols
treated as literals or as data in NOP transformed binaries remains
sound surprisingly often.  Second, the AFL test is stricter because
the rewritten binary must successfully interact with the AFL++ harness
to record a successful execution.

Every rewriter included in this evaluation except multiverse provides
some out-of-the-box support for AFL++ instrumentation.  ddisasm,
retrowrite, and uroboros all produce assembly-level IR that an AFL++
provided tool can directly instrument. Similarly mctoll, reopt, and
revng produce LLVM IR that an AFL++ provided LLVM pass can directly
instrument. E9patch, egalito, and zipr ship with an AFL++
instrumentation pass compatible with their frameworks.

Despite this broad support, only ddisasm, mctoll, retrowrite, and zipr successfully transform 
{\em any} of our test binaries for use with AFL++. In Egalito's case, the included
AFL++ transform requires a patched {\tt afl-fuzz} program \cite{egalito-afl-readme, egalito-afl-readelf, egalito-afl-setup}.
Reopt and revng appear to produce LLVM IR that is not suitable for
transformation. Further, it appears reopt may only perform well on the
NOP transform because it falls back to directly re-linking
sections of the original binary when rewriting fails. Uroboros
appears to fail to produce any functional AFL transformed binaries not
due to any uniform systematic reason but simply because the rewritten
assembly code is very brittle due to incorrect analyses during lifting.

\subsection{Rewriting Success Rates by Compiler}

In this section we present the success rates of our binary rewriters
broken out by the compiler used to generate variants. Binary rewriting success 
is often dependent on the compiler used to produce the input binary as 
many of the heuristics baked into rewriting tools target binary code generation 
logic or optimizations specific to certain compilers. For example, the Intel compiler 
({\tt icx}) in-lines data into the code section on Linux whereas Clang and GCC do not.
As a result of this behavior and other ICX-specific optimizations, some binary rewriters 
have a significantly lower success rate against ICX-produced binaries.

The results restricted to GCC-compiled variants in \autoref{tab:gcc} are most similar to
the aggregate results.  This is unsurprising as GCC is the
prototypical compiler for Linux systems and represents a middle ground
in optimization aggressiveness between the relatively conservative
Clang and the very aggressive Intel ICX. Interestingly, the success rate for Clang-compiled variants (\autoref{tab:gcc}) 
across all tools is slightly higher than GCC's success rate. This could be due to a number
of factors including GCC-only optimizations that prove difficult for binary rewriting tools or GCC
leveraging non-standard ELF file format extensions that are not
produced by Clang.

Our ICX-compiled results are shown in \autoref{tab:ollvm}. They vary widely from 
both GCC and Clang and also across our evaluated tools. DDisasm performs better on
ICX binaries generating an IR in 99\% of cases and generating
functional AFL rewrites 2\% more frequently with these binaries than
in aggregate. By contrast, Egalito's NOP transform success rate drops
for ICX-produced variants, mctoll, Multiverse, and Uroboros are unable to process any
ICX binaries, and Retrowrite and Zipr both perform significantly
worse on ICX binaries although they are still able to successfully
generate functional AFL++ instrumented binaries in some cases.

Given the obfuscation techniques employed by ollvm are meant to hinder binary analysis and reverse engineering, 
we expected the tools to perform worst against ollvm compiled
binaries (\autoref{tab:ollvm}). However, we were surprised to find that in most cases the
rewriting success rate increased for these binaries. It is not clear
if this is because those programs which could be compiled with ollvm
represent the simpler end of our benchmark set, or if there is
something about the ollvm transformations that are amenable to binary
rewriting if not to traditional reverse engineering.

\begin{table*}[!ht]
\centering
\caption{Number and percentage of 1280 GCC-compiled and 1176 Clang-compiled x86-64 Linux binaries for
which the rewriter successfully reaches task checkpoints and passes functional tests.}
  \label{tab:gcc}
  {\rowcolors{4}{lightgray!50}{white}
\begin{tabular}{|c|c|c|c|c|c||c|c|c|c|c|c|}
\cline{2-11}
\multicolumn{1}{c|}{}& \multicolumn{5}{c||}{\textbf{GCC}} & \multicolumn{5}{c|}{\textbf{Clang}} \\
\hline
Tool       & IR      & EXE      & \makecell{Null \\Func.}   & AFL   EXE          & \makecell{AFL \\         Func.}     & IR       & EXE     & \makecell{Null \\Func.} & AFL EXE & \makecell{AFL \\ Func.} \\
\hline
ddisasm    & 1237    & 1146    & 1106    & 1157    & 871   & 1167    & 1034    & 990     & 1057    & 799     \\
\%         & 96.64\% & 89.53\% & 86.40\% & 90.39\% & 68.04\% & 100\% & 87.93\% & 84.18\% & 89.88\% & 67.94\% \\
e9patch    & NA      & 1280    & 994     & 1280    & 453   & NA      & 1174    & 952     & 1174    & 442     \\
\%         & NA      & 100\%   & 77.65\% & 100\%   & 35.39\% & NA    & 99.83\% & 80.95\% & 99.83\% & 37.59\% \\
egalito    & NA      & 1257    & 235     & 1113    & 0     & NA      & 1161    & 494     & 1052    & 0       \\
\%         & NA      & 98.20\% & 18.36\% & 86.95\% & 0     & NA      & 98.72\% & 42.01\% & 89.46\% & 0       \\
mctoll     & 16      & 16      & 16      & 16      & 16    & 11      & 11      & 11      & 11      & 11      \\
\%         & 1.25\%  & 1.25\%  & 1.25\%  & 1.25\%  & 1.25\% & .94\%   & .94\%   & .94\%   & .94\%   & .94\%   \\
multiverse & NA      & 438     & 176     & 0       & 0     & NA      & 442     & 186     & 0       & 0       \\
\%         & NA      & 34.22\% & 13.75\% & 0       & 0     & NA      & 37.59\% & 15.82\%  & 0       & 0       \\
reopt      & 1168    & 874     & 387     & 0       & 0     & 1057    & 901     & 463     & 0       & 0       \\
\%         & 91.25\% & 68.28\% & 30.23\% & 0       & 0     & 89.88\% & 76.62\% & 39.37\% & 0       & 0       \\
retrowrite & 308     & 129     & 124     & 126     & 99    & 288     & 147     & 136     & 146     & 108     \\
\%         & 24.06\% & 10.08\% & 9.69\%  & 9.84\%  & 7.73\% & 24.49\%  & 12.50\%  & 11.56\%  & 12.41\%  & 9.18\%  \\
revng      & NA      & 394     & 354     & 0       & 0     & NA      & 379     & 340     & 0       & 0       \\
\%         & NA      & 30.78\% & 27.66\% & 0       & 0     & NA      & 32.23\% & 28.91\% & 0       & 0       \\
uroboros   & 150     & 92      & 16      & 92      & 0     & 144     & 92      & 56      & 86      & 0       \\
\%         & 11.72\% & 7.19\%  & 1.25\%  & 7.19\%  & 0     & 12.24\%  & 7.82\%  & 4.76\%  & 7.31\%  & 0       \\
zipr       & NA      & 1280    & 1280    & 1033    & 674   & NA      & 1174    & 1174    & 1010    & 671     \\
\%         & NA      & 100\% & 100\% & 80.70\% & 52.66\% & NA      & 99.83\% & 99.83\% & 85.88\% & 57.06\% \\
\hline
\end{tabular}
}

\end{table*}

\begin{table*}[!ht]
\centering
\caption{Number and percentage of 646 ICX-compiled and 244 Ollvm compiled and obfuscated x86-64 Linux binaries for
which the rewriter successfully reaches task checkpoints and passes functional tests.}
  \label{tab:ollvm}
  {\rowcolors{4}{lightgray!50}{white}
\begin{tabular}{|c|c|c|c|c|c||c|c|c|c|c|}
\cline{2-11}
\multicolumn{1}{c|}{}& \multicolumn{5}{c||}{\textbf{ICX}} & \multicolumn{5}{c|}{\textbf{OLLVM}} \\
\hline
Tool       & IR      & EXE     & \makecell{Null \\Func.}     & AFL     EXE & \makecell{AFL \\ Func.} & IR      & EXE     & \makecell{Null \\Func.}     & AFL     EXE & \makecell{AFL \\ Func.} \\
\hline
ddisasm    & 641     & 561     & 552     & 574     & 468   & 237    & 231    & 225    & 232    & 208    \\
\%         & 99.23\% & 86.84\% & 85.44\% & 88.85\% & 72.45\% & 97.13\% & 94.67\% & 92.21\% & 95.08\% & 85.25\% \\
e9patch    & NA      & 646     & 466     & 646     & 214   & NA     & 244    & 208    & 244    & 103    \\
\%         & NA      & 100\% & 72.14\% & 100\% & 33.13\%   & NA     & 100\% & 85.24\% & 100\% & 42.21\% \\
egalito    & NA      & 636     & 136     & 119     & 0     & NA     & 240    & 118    & 209    & 0      \\
\%         & NA      & 98.45\% & 21.05\%  & 18.42\% & 0     & NA     & 98.36\% & 48.36\% & 85.66\% & 0      \\
mctoll     & 0       & 0       & 0       & 0       & 0     & 3      & 3      & 3      & 3      & 3      \\
\%         & 0       & 0       & 0       & 0       & 0     & 1.23\%  & 1.23\%  & 1.23\%  & 1.23\%  & 1.23\%  \\
multiverse & NA      & 0       & 0       & 0       & 0     & NA     & 0      & 0      & 0      & 0      \\
\%         & NA      & 0       & 0       & 0       & 0     & NA     & 0      & 0      & 0      & 0      \\
reopt      & 575     & 571     & 173     & 0       & 0     & 207    & 210    & 111    & 0      & 0      \\
\%         & 89.01\% & 88.39\% & 26.78\%  & 0       & 0     & 84.84\% & 86.07\% & 45.49\% & 0      & 0      \\
retrowrite & 157     & 24      & 23      & 24      & 18    & 64     & 34     & 26     & 34     & 29     \\
\%         & 24.30\% & 3.71\%  & 3.56\%  & 3.71\%  & 2.79\% & 26.23\% & 13.93\% & 10.66\%  & 13.93\% & 11.89\%  \\
revng      & NA      & 24      & 16      & 0       & 0     & NA     & 88     & 76     & 0      & 0      \\
\%         & NA      & 3.71\%   & 2.47\%   & 0       & 0     & NA   & 36.07\% & 31.15\% & 0      & 0      \\
uroboros   & 4       & 0       & 0       & 0       & 0     & 66     & 32     & 24     & 32     & 0      \\
\%         & .62\%   & 0       & 0       & 0       & 0     & 27.05\% & 13.11\%  & 9.84\%  & 13.11\%  & 0      \\
zipr       & NA      & 646     & 646     & 479     & 154   & NA     & 244    & 244    & 186    & 110    \\
\%         & NA      & 100\% & 100\% & 74.15\% & 28.84\%  & NA     & 100\% & 100\% & 76.23\% & 45.08\% \\
\hline
\end{tabular}
}

\end{table*}

\subsection{Analysis of Binary Rewriter Success}

In this section, we investigate binary formatting options to determine if they are 
correlates for binary rewriter success. We collected three readily-identifiable formatting features using {\tt readelf} from GNU {\tt binutils}: (1) whether or not the binary is position-independent, (2) whether or not the binary is stripped, and (3) 
the sections included in the binary\footnote{We exclude sections which appear in all binaries 
(e.g., .text) and sections unique to specific program (e.g., .gresource.gnome\_calculator).}.

We collated the success and failure rate across these features for each tool against our
corpus of variants considering both rewriting tasks (i.e., NOP and AFL).  Then we 
identified the four most predictive features for
rewriting success or failure of the AFL transform for each tool. These features are
presented in \autoref{tab:tree}.  In many cases these features are
expected and match the advertised capabilities of each tool. For
example, retrowrite only supports relocatable (i.e., {\tt pi}) and non-stripped (i.e., {\tt strip})
binaries which are its two most predictive features for success.

Next we train a decision tree based on this feature collection to
predict the likelihood of success of each rewriter against an example
binary when using the AFL transform. Before training we use linear
support vector classification to select the most discriminating
features for that rewriter. The resulting decision trees are printed
as Python code in Appendix \ref{trees}.  We evaluate the resulting decision
tree using 70\% of our binaries for training and reserving 30\% for
testing.  The accuracy of the resulting tree is shown in
\autoref{tab:tree}.

\begin{table*}[!htb]
  \centering
  \caption{Decision tree accuracy predicting binary rewriting success
    based on simple binary features}
  \label{tab:tree}
  {\rowcolors{2}{lightgray!50}{white}
 \begin{tabular}{|r|r|l|l|}
 \hline
 Rewriter   & NOP     & AFL    & Most Predictive Features (AFL) \\
 \hline
 ddisasm    & 90.03\% & 81.47\%  & \verb|note.abi-tag|, \verb|interp|, \verb|gcc_except_table| \verb|debug_str|\\
 e9patch    & 80.57\% & 86.06\% & \verb|pi|, \verb|got.plt|, \verb|data.rel.ro|, \verb|plt.got| \\
 egalito    & 87.15\% & &        \\
 mctoll     & 98.80\% & 98.80\% & \verb|strip|, \verb|data.rel.ro|, \verb|symtab|, \verb|strtab|\\
 multiverse & 97.80\% &  &       \\
 reopt      & 67.82\% &  &       \\
 retrowrite & 94.32\% & 93.02\% & \verb|pi|, \verb|strip|, \verb|symtab|, \verb|strtab| \\
 revng      & 78.78\% & & \\
 uroboros   & 96.31\% &  &       \\
 zipr       & 86.65\% & 79.98\% & \verb|strip|, \verb|note.gnu.build-id|, \verb|symtab|, \verb|strtab|\\
 \hline
 \end{tabular}
}

\end{table*}

As shown in \autoref{tab:tree} the resulting decision trees, despite
their reliance on very simple binary features were very accurate in
predicting the chances of tool success.  We anticipate two benefits
from this analysis.  First, tool developers will have insight into
properties of binaries that cause their rewriting tools to fail.
Second, users can nearly instantaneously run a
combination of {\tt readelf} and our decision tree to see what tools,
if any, will reliably transform a given target binary.  This is useful
when many binary rewriting tools can run for minutes and even hours on
a single binary.  The success of this simple machine learning model trained on
simple inputs indicates promising new directions for the practical
application of binary rewriting technology discussed in Section
\ref{sec:conclusion}. The decision trees and the code used to build and train 
them are included in our publicly available
artifact repository. 


\subsection{Size of Rewritten Binaries}
\label{subsec:size}


Changes in binary size (shown in column 3 of \autoref{tab:perf}) reflect a tool's 
design decisions and can impact the 
utility, efficiency, and potential use cases for the
rewritten binary. On average, ddisasm and retrowrite's rewritten binaries are slightly smaller, likely due to symbol and debug information dropped during their rewriting processes.
For Egalito, mctoll, uroboros, and zipr the size of the binary increases by a non-trivial amount. 
Revng and Multiverse are outliers with rewritten binaries that are nearly 16 and 9
times the size of the original, respectively. For Revng, the recompiled binaries exhibit very large increases in unmapped areas of the binary (223 times on average!), potentially indicating flaws in the recompilation stage that can be rectified. For Multiverse the increase is due to a defining design
decision: it produces rewritten binaries that contain all {\em possible}
disassemblies of the original binary.

It is important to note that size increases are calculated for binaries the tool can successfully
rewrite, which varies for each tool. As a result, a direct comparison of tools using \autoref{tab:perf} is not possible.
As such, we also conducted a comparative evaluation of binary size increases between pairs of tools where each successfully rewrote one or more of the same programs (data shown in \autoref{supptables}). The best performing tools were retrowrite and ddisasm which produced binaries that were on average  62\% and 65\% the size of those produced by the other tools, respectively. The tools that produced the largest binaries were revng and multiverse, which produced binaries that were approximately 13 and 7 times the size of those produced by the other tools, respectively.

To investigate these size changes further, we 
analyzed size changes per section per rewriting tool (data shown in \autoref{supptables}). 
Note that in many cases rewriting tools break elf section tables. In these cases bloaty~\cite{bloaty}, the
tool we use to collect section size, is unable to determine sizes for
that section in corresponding binaries. In nearly every rewriting tool
the largest increase in size of the elf file is in unmapped bytes or
bytes that are not accounted for by the section table.  This is likely
due to at least the following two factors.  First, because any extra
non-standard runtime harnesses or extra rewriting-specific supports
are not properly entered into the section table of the rewritten
binary.  Second, binary rewriting tools are not penalized for dropping
sections or breaking parts of the section header table that are not
required for execution.

\subsection{Binary Rewriter Performance}
To accurately and successfully rewrite a binary executable requires
significant static analysis.  These analyses often scale
super-linearly with the size of the program being rewritten.  We
summarize the average run time of each tool in \autoref{tab:perf}.

\begin{table}[!htb]
\centering
\caption{Average tool runtime memory high-water mark, and relative program size change across successful rewrites}
\label{tab:perf}
{\rowcolors{2}{lightgray!50}{white}
\begin{tabular}{|c|c|c|c|}
\hline
           & Runtime   & Memory high-   & Relative program\\
Tool       & (seconds) & water (kbytes) & size change\\
\hline
ddisasm    & 72.81   & 509215.92   & 91.90\%   \\
e9patch    & 2.74    & 105132.97   & 114.45\%  \\
egalito    & 454.40  & 10433233.07 & 169.17\%  \\
mctoll     & 0.00    & 1405.30     & 128.22\%  \\
multiverse & 1195.72 & 687609.83   & 870.71\%  \\
reopt      & 169.89  & 4061695.00  & 99.61\%   \\
retrowrite & 114.57  & 1967393.19  & 83.78\%   \\
revng      & 703.74  & 2244106.85  & 1581.95\% \\
uroboros   & 19.17   & 93575.58    & 148.97\%  \\
zipr       & 233.61  & 1015891.15  & 140.05\%  \\
\hline
\end{tabular}
}

\end{table}

As with rewritten program size, the reported averages are skewed
because they are calculated across the set of binaries successfully
rewritten by each tool.  Thus, rewriters that successfully rewrite
larger and more complicated binaries have an average that skews
higher.  To account for this we also present the comparative average
in \autoref{tab:comp-time}. In each cell, the comparative average tool
runtime across successfully rewritten binaries by both tools is
expressed as a percentage of the row tool to the column tool. For
example, uroboros runtime is roughly 14.49\% of the runtime of ddisasm.
A significant trend observable in \autoref{tab:comp-time} is that
tools with higher success rates tend to run longer than tools with low
rewriting success rates.  This is not surprising as successful tools
perform more analyses and more detailed analyses.  They also
explicitly handle more portions of the ELF file and more edge cases.

\begin{table*}[!htb]
\centering
\caption{Comparative runtime of binary rewriting tools}
\label{tab:comp-time}
{\rowcolors{2}{lightgray!50}{white}
\begin{tabular}{|c|c|c|c|c|c|c|c|c|c|c|}
\hline
Tool       & ddisasm   & e9patch    & egalito  & mctoll         & multiverse & reopt    & retrowrite & revng    & uroboros   & zipr  \\
\hline
ddisasm    & 100\%     & 2656.96\%  & 15.95\%  & 35288840.58\%  & 4.40\%  & 42.86\%  & 26.97\%   & 10.35\%  & 689.90\%   & 33.68\%   \\
e9patch    & 3.76\%    & 100\%      & 0.59\%   & 1328163.77\%   & 0.19\%  & 1.61\%   & 2.23\%    & 0.39\%   & 15.49\%    & 1.20\%    \\
egalito    & 626.77\%  & 16963.74\% & 100\%    & 219758856.52\% & 18.30\% & 270.91\% & 503.69\%  & 64.57\%  & 2123.71\%  & 205.48\%  \\
mctoll     & 0.00\%    & 0.01\%     & 0.00\%   & 100\%          & 0.00\%  & 0.00\%   & 0.00\%    & 0.00\%   & 0.00\%     & 0.00\%    \\
multiverse & 2270.67\% & 51759.60\% & 546.47\% & 441563025.00\% & 100\%   & 749.41\% & 1983.34\% & 224.58\% & 20301.47\% & 1550.43\% \\
reopt      & 233.32\%  & 6199.21\%  & 36.91\%  & 82335688.41\%  & 13.34\% & 100\%    & 166.07\%  & 24.14\%  & 811.10\%   & 74.87\%   \\
retrowrite & 370.73\%  & 4485.01\%  & 19.85\%  & 32450926.47\%  & 5.04\%  & 60.22\%  & 100\%     & 19.32\%  & 33.45\%    & 50.96\%   \\
revng      & 966.48\%  & 25678.91\% & 154.87\% & 341058002.90\% & 44.53\% & 414.23\% & 517.54\%  & 100\%    & 4073.82\%  & 300.14\%  \\
uroboros   & 14.49\%   & 645.59\%   & 4.71\%   & 7081250.00\%   & 0.49\%  & 12.33\%  & 298.95\%  & 2.45\%   & 100\%      & 10.68\%   \\
zipr       & 296.89\%  & 8349.20\%  & 48.67\%  & 104211334.78\% & 6.45\%  & 133.56\% & 196.23\%  & 33.32\%  & 936.69\%   & 100\%     \\
\hline
\end{tabular}
}

\end{table*}

\subsection{Binary Rewriter Memory High-Water Mark}
Like runtime, a tool's memory requirements may make rewriting impractical.  For
large binaries, memory requirements will frequently outstrip the
memory available on a server class machine.  We present the
average memory high-water mark during binary rewriting in
column 2 of \autoref{tab:perf} and for the same reasons described in \autoref{subsec:size} we also conducted a
comparative evaluation of memory high-water marks (data shown in \autoref{supptables}). As with rewriter performance,
we generally find that successful tools require more resources during rewriting.

\subsection{Functionality Against Full Test Suite}
For three benchmark programs with readily available test suites we
check the degree to which successful execution of the Null Function test
predicts successful execution of the complete test suite\footnote{We 
report full results for only three benchmark programs due to
the dearth of high-quality test suites for real-world programs and the
high level of effort required for properly configuring them.}.
Our results are presented in \autoref{tab:test-func}. In each cell we report
the number of binaries that completely pass the full test suite and
the number of binaries that pass the Null Function test as
``full/null''.  We report this for each binary rewriter as well as
for the original input binaries files. There are up to 60 binaries for each
program due to the multiple build options (e.g., compiler, optimization level, pie,
stripped, etc.). We do not include ICX-compiled binaries due to the extra runtime dependencies 
they require that impose significant extra burden when running full test suites in the 
test environment.

\begin{table}[!htb]
\centering
\caption{Number of binaries passing their full test suite versus passing the Null Function test  }
\label{tab:test-func}
{\rowcolors{2}{lightgray!50}{white}
\begin{tabular}{|c|c|c|c|}
\hline
Tool       & lighttpd & nginx & redis \\
\hline
original   &   30/30  & 60/60 & 26/30 \\
ddisasm    &    0/30  & 60/60 & 26/30 \\
e9patch    &   30/30  & 60/60 & 26/30 \\
egalito    &   18/18  & 18/18 & 10/10 \\
multiverse &    0/0   &  0/0  &  0/0  \\
reopt      &    2/19  &  4/60 &  2/8  \\
retrowrite &    0/9   & 16/26 &  0/0  \\
revng      &    0/0   &  0/0  &  0/0  \\
uroboros   &    0/0   &  0/0  &  0/0  \\
zipr       &   28/30  & 58/58 & 22/30 \\
\hline
\end{tabular}
}

\end{table}

Overall we find a weak correlation with only 258 rewritten programs passing their full test suite
of the 395 rewritten programs that passed their Null Function tests.
However, it is worth noting that some original binaries (i.e., inputs to the binary rewriter) that pass
the Null Function test do not pass the full test suite (e.g., {\tt redis} when
compiled with {\tt -Ofast}).

Note that we disable one test in {\tt redis} because it looks for a specific
symbol in the stack trace. This is sufficiently {\em internal}
that we believe it does not compromise program soundness for binary rewriters to change this
behavior.

\subsection{Performance Against Full Test Suite}
The performance of rewritten binaries is critical to many use cases
for static binary rewriting. If performance degradation exceeds that
of dynamic binary rewriting then dynamic rewriting is often a better
alternative as it is able to leverage runtime information to more
reliably transform program behavior.  We report the change in runtime
and memory requirements for successfully rewritten programs running
against their full test suite in \autoref{tab:test-perf}. Only those
rewriters which produced binaries capable of passing all tests are
included.  With the exception of Reopt-rewritten binaries which had
resource consumption at least an order of magnitude over the original,
runtime and memory consumption of the rewritten binaries is close to
that of the original binary.  This is especially true of the memory
high-water mark.

\begin{table}[!htb]
\centering
\caption{Average performance of rewritten binaries when run against the full test suite}
\label{tab:test-perf}
{\rowcolors{2}{lightgray!50}{white}
\begin{tabular}{|c|r|r|}
\hline
           & Runtime   & Memory High-water \\
Tool       & (seconds) & Mark (kbytes)     \\
\hline
ddisasm    & 109.43\%  & 100.21\%        \\
e9patch    & 119.53\%  & 99.06\%         \\
egalito    & 104.45\%  & 99.85\%         \\
reopt      & 1324.66\% & 51937.17\%      \\
retrowrite & 103.84\%  & 100.17\%        \\
zipr       & 102.50\%  & 102.38\%        \\
\hline
\end{tabular}
}

\end{table}
\section{Discussion}
\label{sec:conclusion}

We identify several trends with respect to binary rewriter IR from our results.
First, rewriting via LLVM IR appears to be infeasible given the current state of
binary type analysis. Only one binary (the trivial {\tt hello-world}) was
successfully instrumented for AFL++ using an LLVM rewriter (mctoll).
Additionally, non-trivial NOP-transformed binaries successfully produced  
with reopt had dramatically increased runtime and memory consumption as compared to
the original. Second, direct rewriting as performed by Egalito and Zipr successfully 
produced executables even in the presence of analysis errors; however their output binaries also 
demonstrated a higher functional failure rate. Conversely, reassemblable disassemblers 
were more likely to raise errors during re-assembling and re-linking and thus fail to create an executable.
Trampoline rewriters such as e9patch are {\em very} reliable across a wide range of binaries
if only additive instrumentation and no modification of existing code is required.

During our evaluation, we communicated with the developers of our 
evaluated tools to share our partial results and our benchmark set.%
\footnote{We did not communicate with the rev.ng developers until {\em
  after} our evaluation.  As a result our evaluation did not include
the recommended but non-default arguments ``{\tt -O2 -i}'' which are
expected to result in reduced rewritten binary code size and runtime
but are not expected to impact rewritten binary functionality}
Unsurprisingly, the best-performing tools in our evaluation, ddisasm
and zipr, had sufficient development resources to respond
to specific failures encountered in our work. Thus, their performance 
against this evaluation set likely outperforms their expected performance
in general. This is indicative of a {\em defining} characteristic of 
binary rewriting at this point in time; binary rewriting is eminently 
practical in many {\em particular} cases that have been addressed and 
considered by tool developers, but impossible in the {\em general} case 
as the universe of binary formats and features is simply too large with 
too many edge cases to handle.

Our evaluation indicates that practical
applications of binary rewriting should be preceded by a scoping stage.
In this stage, the target binary is classified as either ``in scope'' 
or ``out of scope'' for the binary rewriting tool(s) of interest.
While scoping can be accomplished via traditional binary analysis, the 
high success rate demonstrated by our simple predictive model with 
trivially collected features shows that accurate scoping can be conducted 
with little effort. Further, the relative simplicity of our model implies 
that more reliable and accurate predictive models are likely easily
within reach. With such a model, users of binary rewriting tools may quickly ensure 
their target binaries meet the expectations of the available rewriting tools
before initiating expensive binary rewriting tasks. For binaries that 
are likely to fail during static rewriting, the user could either conserve 
their resources by forgoing binary rewriting or spend them employing more 
expensive techniques such as dynamic binary rewriting.

\section{Conclusion} 

In this work, we evaluated and compared ten binary 
rewriting tools on two rewriting tasks across a corpus of 3344 
variant binaries produced using three compilers and 34 benchmark 
programs. Our evaluation measured the performance of the tools 
themselves as well as the performance and soundness of the rewritten
binaries they produce. In general, our evaluation indicates that 
binary rewriters that lift to high-level machine-independent IRs 
(e.g., LLVM IR) were much less successful in terms of generality and 
reliability. Additionally, we identified binary features that 
are predictive of rewriting success and showed that a simple
decision tree model trained on these features can accurately predict 
whether a particular tool can rewrite a target binary. The findings 
and artifacts contributed by this evaluation have been made publicly available and are intended to support
users and developers of binary rewriting tools and drive rewriter 
adoption and maturation.

\section*{Artifact Availability}

We have made the full set of artifacts generated in this work
including our evaluation infrastructure, corpus of test binaries,
predictive models, and the evaluated tools publicly available at:

\begin{center}\url{https://gitlab.com/GrammaTech/lifter-eval} \cite{lifter-eval}\end{center}

\begin{acks}
This material is based upon work supported by the Office of Naval Research (ONR) under Contract No. N00014-21-C-1032.  Any opinions, findings and conclusions or recommendations expressed in this material are those of the author(s) and do not necessarily reflect the views of the ONR.
\end{acks}

\bibliographystyle{ACM-Reference-Format}
\bibliography{lifter-eval.bib}

\newpage
\onecolumn
\appendix

\section{Supplementary Performance Tables}
\label{supptables}

\begin{table*}[!htb]
\centering
\caption{Comparative size of rewritten binaries in the
intersection of those programs which are successfully rewritten by both tools. The percentage of the successfully rewritten binary sizes 
by both tools are calculated as a ratio of the row tool to the column tool. For example, multiverse rewritten
binaries are just over 9 times bigger on average than ddisasm rewritten binaries.
An entry of ``NA'' indicates that {\em no} binaries were successfully
rewritten by both tools.
}
\label{tab:comp-size}
{\rowcolors{2}{lightgray!50}{white}
\begin{tabular}{|c|c|c|c|c|c|c|c|c|c|c|c|c|}
\hline
Tool       & ddisasm   & e9patch   & egalito  & mctoll    & multiverse & reopt     & retrowrite & revng   & uroboros  & zipr     \\
\hline
ddisasm    & 100\%     & 79.65\%   & 51.45\%  & 100.52\%  & 10.67\%    & 92.59\%   & 106.39\%   & 6.47\%  & 64.51\%   & 63.82\%  \\
e9patch    & 125.54\%  & 100\%     & 67.71\%  & 98.09\%   & 13.06\%    & 115.77\%  & 141.65\%   & 7.59\%  & 84.26\%   & 81.72\%  \\
egalito    & 194.36\%  & 147.69\%  & 100\%    & 189.77\%  & 21.58\%    & 167.93\%  & 164.72\%   & 10.54\% & 95.58\%   & 121.13\% \\
mctoll     & 99.48\%   & 101.94\%  & 52.69\%  & 100\%     & 27.08\%    & 351.83\%  & 98.89\%    & 3.54\%  & 129.17\%  & 84.60\%  \\
multiverse & 937.59\%  & 765.59\%  & 463.33\% & 369.23\%  & 100\%      & 854.06\%  & 907.58\%   & 51.55\% & 567.15\%  & 626.59\% \\
reopt      & 108.00\%  & 86.38\%   & 59.55\%  & 28.42\%   & 11.71\%    & 100\%     & 114.93\%   & 5.89\%  & 56.22\%   & 70.62\%  \\
retrowrite & 93.99\%   & 70.60\%   & 60.71\%  & 101.12\%  & 11.02\%    & 87.01\%   & 100\%      & 6.62\%  & NA       & 67.52\%  \\
revng      & 1546.44\% & 1318.30\% & 949.20\% & 2823.92\% & 193.97\%   & 1696.73\% & 1509.82\%  & 100\%   & 1463.48\% & 982.19\% \\
uroboros   & 155.01\%  & 118.68\%  & 104.63\% & 77.42\%   & 17.63\%    & 177.89\%  & NA        & 6.83\%  & 100\%     & 109.41\% \\
zipr       & 156.68\%  & 122.37\%  & 82.56\%  & 118.20\%  & 15.96\%    & 141.60\%  & 148.11\%   & 10.18\% & 91.40\%   & 100\%    \\
\hline
\end{tabular}
}

\end{table*}

\begin{table*}[!htb]
  \centering
  \caption{Average size change by section for each rewriting tool}
  \label{tab:section-sizes}
  {\rowcolors{2}{lightgray!50}{white}
\begin{tabular}{|c|r|r|r|r|r|r|r|r|r|r|}
\hline
Section                   & ddisasm  & e9patch    & egalito  & mctoll   & multiverse & reopt    & retrowrite & revng      & uroboros  & zipr     \\
\hline
.got.plt                  & 100.08\% & 100.00\%   & 100.67\% & 89.60\%  & 100.00\%   & 100.29\% & 100.00\%   & NA         & 99.24\%   & NA       \\
.data                     & 102.94\% & 100.00\%   & 100.24\% & NA       & 99.99\%    & 103.29\% & 100.42\%   & 1014.94\%  & 101.87\%  & NA       \\
.dynamic                  & 97.48\%  & 100.00\%   & 66.53\%  & 97.70\%  & 100.00\%   & 99.82\%  & 101.83\%   & NA         & 99.91\%   & NA       \\
.rela.dyn                 & 87.27\%  & 114.67\%   & 823.91\% & 54.35\%  & 100.00\%   & 102.65\% & 97.30\%    & 355.43\%   & 91.19\%   & NA       \\
.strtab                   & 104.90\% & 100.00\%   & 77.54\%  & 95.27\%  & 99.26\%    & 76.69\%  & 105.45\%   & 407.41\%   & 485.39\%  & NA       \\
.dynsym                   & 75.44\%  & 100.00\%   & 101.86\% & 75.24\%  & 100.00\%   & 110.20\% & 77.96\%    & 821.27\%   & 99.43\%   & NA       \\
.dynstr                   & 72.78\%  & 100.00\%   & 101.42\% & 73.18\%  & 99.99\%    & 104.62\% & 74.56\%    & 1075.44\%  & 99.75\%   & NA       \\
.symtab                   & 118.87\% & 100.00\%   & 84.73\%  & 93.93\%  & 100.00\%   & 107.44\% & 95.09\%    & 270.77\%   & 1675.22\% & NA       \\
.eh\_frame\_hdr           & 103.76\% & 100.00\%   & NA       & 110.43\% & 100.00\%   & 106.36\% & 93.36\%    & 25.33\%    & 108.05\%  & NA       \\
.plt                      & 100.05\% & 100.00\%   & 100.45\% & 89.60\%  & 100.00\%   & 100.28\% & 99.81\%    & 990.99\%   & 99.75\%   & NA       \\
.rela.plt                 & 99.93\%  & 100.00\%   & 99.35\%  & NA       & 100.00\%   & 100.30\% & 99.81\%    & 757.03\%   & 100.00\%  & NA       \\
.eh\_frame                & 109.75\% & 100.00\%   & NA       & 101.88\% & 100.00\%   & 125.52\% & 15.14\%    & 520.42\%   & 111.34\%  & NA       \\
{[}ELF Program Headers{]} & 96.37\%  & 100.00\%   & 94.29\%  & 105.53\% & 144.44\%   & 94.00\%  & 104.79\%   & NA         & 97.77\%   & 119.81\% \\
{[}ELF Section Headers{]} & 97.66\%  & 100.00\%   & 71.70\%  & 94.93\%  & 116.13\%   & 103.62\% & 92.86\%    & 148.37\%   & 97.57\%   & 98.55\%  \\
.rodata                   & 100.10\% & 100.00\%   & 100.43\% & NA       & 100.00\%   & 99.96\%  & 100.06\%   & 480.80\%   & 100.03\%  & NA       \\
.text                     & 146.85\% & 100.00\%   & 161.64\% & 100.89\% & 100.00\%   & 240.13\% & 120.45\%   & 3817.02\%  & 110.17\%  & NA       \\
{[}Unmapped{]}            & 128.93\% & 13162.97\% & 670.47\% & 350.40\% & 2285.56\%  & 181.98\% & 225.89\%   & 22384.87\% & 165.31\%  & 10.40\%  \\
\hline
\end{tabular}
}

\end{table*}

\begin{table*}[!htb]
\centering
\caption{Comparative memory high-water mark in kilobytes between
rewriting tools. The comparative memory high-water mark across successfully rewritten binaries
by both tools is expressed as a percentage of the row tool to 
the column tool. For example, uroboros' maximum memory consumption is roughly 15.14\% of the 
maximum memory consumption of ddisasm.} 
\label{tab:comp-memory}
{\rowcolors{2}{lightgray!50}{white}
\begin{tabular}{|c|c|c|c|c|c|c|c|c|c|c|}
\hline
Tool       & ddisasm   & 39patch    & egalito & mctoll      & multiverse & reopt    & retrowrite & revng  & uroboros   & zipr      \\
\hline
ddisasm    & 100\%     & 484.35\%   & 4.69\%  & 36235.17\%  & 76.88\%   & 12.54\%  & 24.70\%  & 22.69\%  & 660.63\%   & 53.68\%   \\
e9patch    & 20.65\%   & 100\%      & 0.99\%  & 7481.13\%   & 13.13\%   & 2.59\%   & 5.39\%   & 4.68\%   & 119.91\%   & 10.57\%   \\
egalito    & 2133.00\% & 10124.69\% & 100\%   & 741944.13\% & 1446.63\% & 259.70\% & 499.62\% & 471.95\% & 11511.47\% & 1048.77\% \\
mctoll     & 0.28\%    & 1.34\%     & 0.01\%  & 100\%       & 0.23\%    & 0.03\%   & 0.09\%   & 0.06\%   & 1.64\%     & 0.14\%    \\
multiverse & 130.07\%  & 761.64\%   & 6.91\%  & 43096.38\%  & 100\%     & 18.42\%  & 66.16\%  & 38.14\%  & 806.39\%   & 111.25\%  \\
reopt      & 797.64\%  & 3863.39\%  & 38.51\% & 289025.13\% & 542.85\%  & 100\%    & 206.36\% & 180.99\% & 4336.39\%  & 407.09\%  \\
retrowrite & 404.93\%  & 1856.78\%  & 20.02\% & 112051.06\% & 151.14\%  & 48.46\%  & 100\%    & 94.37\%  & 251.50\%   & 197.87\%  \\
revng      & 440.70\%  & 2134.54\%  & 21.19\% & 159687.83\% & 262.22\%  & 55.25\%  & 105.97\% & 100\%    & 2492.68\%  & 229.02\%  \\
uroboros   & 15.14\%   & 83.40\%    & 0.87\%  & 6083.90\%   & 12.40\%   & 2.31\%   & 39.76\%  & 4.01\%   & 100\%      & 11.28\%   \\
zipr       & 186.28\%  & 945.96\%   & 9.53\%  & 70240.87\%  & 89.89\%   & 24.56\%  & 50.54\%  & 43.66\%  & 886.33\%   & 100\%     \\
\hline
\end{tabular}
}

\end{table*}

\newpage
\section{Predictive Models}
\label{trees}

\begin{figure}[!htb]
  \caption{Decision tree to predict the success of AFL instrumentation with ddisasm.  Accuracy of 81.47\%.}
  \label{lst:ddisasm-tree}
  \adjustbox{max size={0.5\textwidth}{0.5\textheight}}{
    \input{Listings/ddisasm}  
  }
\end{figure}

\begin{figure}[!htb]
  \caption{Decision tree to predict the success of AFL instrumentation with e9patch.  Accuracy of 86.06\%.}
  \label{lst:e9patch-tree}
  \adjustbox{max size={0.5\textwidth}{0.5\textheight}}{
    \input{Listings/e9patch}  
  }
\end{figure}

\begin{figure}[!htb]
  \caption{Decision tree to predict the success of AFL instrumentation with mctoll.  Accuracy of 98.80\%.}
  \label{lst:mctoll-tree}
  \adjustbox{max size={0.5\textwidth}{0.5\textheight}}{
    \input{Listings/mctoll}  
  }
\end{figure}

\begin{figure}[!htb]
  \caption{Decision tree to predict the success of AFL instrumentation with retrowrite.  Accuracy of 93.02\%.}
  \label{lst:retrowrite-tree}
  \adjustbox{max size={0.5\textwidth}{0.5\textheight}}{
    \input{Listings/retrowrite}  
  }
\end{figure}

\begin{figure}[!htb]
  \caption{Decision tree to predict the success of AFL instrumentation with zipr.  Accuracy of 79.98\%.}
  \label{lst:zipr-tree}
  \adjustbox{max size={0.5\textwidth}{0.5\textheight}}{
    \input{Listings/zipr}
  }
\end{figure}

\end{document}